# A multimodal laser-scanning nonlinear optical microscope with a rapid broadband Fourier-transform coherent Raman modality

Faris Sinjab[1], Kazuki Hashimoto[2], Venktata Ramaiah Badarla[1], Junko Omachi[1,3], and Takuro Ideguchi[1,2,4,*]

[1] Institute for Photon Science and Technology, The University of Tokyo, Tokyo, 113-0033, Japan

[2] Department of Physics, University of Tokyo, Tokyo, 113-0033, Japan

[3] Current address: School of Science and Technology, Kwansei Gakuin University, Sanda, Hyogo 669-1337, Japan

[4] PRESTO, Japan Science and Technology Agency, Tokyo 113-0033, Japan

[*] ideguchi@ipst.s.u-tokyo.ac.jp

**Abstract**

Nonlinear optical microscopy allows rapid high-resolution microscopy with image contrast generated from intrinsic properties of the sample. Established modalities such as multiphoton excited fluorescence and second/third-harmonic generation can be combined with other nonlinear techniques, such as coherent Raman spectroscopy which typically allow chemical imaging of a single resonant vibrational mode of a sample. Here, we utilize a single ultrafast laser source to obtain broadband coherent Raman spectra on a microscope, together with other nonlinear microscopy approaches on the same instrument. We demonstrate that the coherent Raman modality allows broadband measurement (>1000 $cm^{-1}$), with high spectral resolution (<5 $cm^{-1}$), with a rapid spectral acquisition rate (3-12 kHz). This enables Raman hyperspectral imaging of > kilo-pixel images at >11 frames per second.

**Introduction**

Nonlinear optical microscopy has enabled a variety of high-resolution imaging techniques with applications across materials and biological sciences [1]. Several nonlinear optical processes excited by near-IR pulsed lasers can be used to generate image contrast, including multiphoton excited fluorescence (MPEF) [2-4], second- and third-harmonic generation (SHG/THG) [4-8], stimulated Raman scattering (SRS) and coherent anti-Stokes Raman scattering (CARS) [9-11].

MPEF can be generated in materials with absorption transitions at multiples of the excitation laser frequency, e.g. with a mode-locked Ti:Sapphire laser centered near 800 nm, two-photon excited fluorescence (2PEF) can occur above 400 nm, and three-photon excited fluorescence (3PEF) above 267 nm can be measured, as shown in Fig. 1(a). SHG at twice the excitation frequency will occur in materials with non-centrosymmetric inversion symmetry [5] and THG is generated at regions of high heterogeneity in the refractive index [6-8].

Coherent Raman spectroscopies such as CARS and SRS are nonlinear optical processes mediated via the vibrational energy levels of the system under investigation [9-11]. Here, the frequency difference between two pulsed lasers is tuned to coherently drive a particular vibrational transition, followed by a probe measurement producing either a photon of a different (CARS) or identical (SRS) optical frequency for detection. Alternatively, using an ultrafast femtosecond (fs) pulse with duration shorter than the vibrational time period allows impulsive SRS excitation, where many vibrational modes can be driven at once [12-14]. The subsequent temporal evolution of the refractive index modulation at the vibrational motion timescale can then be probed with an ultrashort probe beam, the blue-shift of which can be detected by optical filtering. Subsequent pump/probe pulse pairs with scanned delay can trace the molecular vibration at a down-converted timescale, dependent on the scanning speed. This down-converted vibrational profile can then be Fourier-transformed to extract the FT-CARS spectrum. Various approaches to FT-CARS spectroscopy have been demonstrated in this way, primarily differing in their mechanism for delay scanning, which sets the spectral acquisition rate and consequently the pixel dwell time for imaging. Using a spatial light modulator in a pulse-shaping configuration, the delay can be controlled electronically, limited by the refresh time of the device (~10 ms) [12]. A dual frequency-comb approach can be used to accurately measure a high-resolution spectrum in a short time (~10-20 µs), but requires a relatively long recycle time related to the repetition rate difference between each comb source (~0.5-10 ms), resulting in an inefficient measurement duty cycle [14-16]. Acousto-optic programmable dispersive filters (AOPDF) have also recently been utilized for micro-spectroscopy of low-wavenumber bands [17,18], but would introduce significant material dispersion for ultrashort (<20 fs) pulses required for broadband spectroscopy. Alternatively, simple mechanical delay scanning in an interferometer is possible, either with a motorized stage [13,19], or with rapid scanning mirrors to enable spectral acquisition with 20.0-333.3 µs recycling time [20,21]. A resonant scanning mirror system was recently demonstrated for hyperspectral CARS microscopy at 2.4 fps of 100×100 pixel images with 37 $cm^{-1}$ spectral resolution [22].

Previous studies have combined one or more of MPEF, SHG, and THG imaging with CARS/SRS microscopy, with the first demonstrations using the standard two-color picosecond pulsed excitation approach for CARS imaging of a single vibrational resonance [23,24]. Following this, several multimodal approaches have extended the Raman modality to include varying levels of hyperspectral imaging. Pope et al. used a single 5 fs laser source for spectral focusing CARS imaging of multiple simultaneous bands tunable from 1200-3800 $cm^{-1}$ (10 $cm^{-1}$ resolution, 0.01 ms pixel dwell time for three simultaneous images at separately tunable bands) with SHG and 2PEF [25]. Camp Jr. et al. demonstrated ultra-broadband CARS (500-3800 $cm^{-1}$, <10 $cm^{-1}$ resolution, 3.5 ms pixel dwell time) using a femtosecond supercontinuum CARS pump source with a picosecond probe laser, which allows for simultaneous SHG and 2PEF [26]. Kumar et al. used a scanning notch filter for pulse shaping of a mode-locked Ti:Sapphire laser to produce CARS spectra (600-1400 $cm^{-1}$, 5 ms pixel dwell time), combined with SHG/THG/2PEF as well as vibrational non-resonant four-wave mixing (FWM) and optical coherence tomography [27]. Yoneyama et al. used a microchip laser source with a photonic crystal fiber to generate a supercontinuum, combined with fast CCD detection for ultra-broadband CARS (600-3200 $cm^{-1}$, <1 $cm^{-1}$ resolution, 0.8 ms pixel dwell time) [28], with a similar system also previously shown to enable third-order sum frequency generation [29]. Time-domain fluorescence spectroscopy combined with FT-CARS has also recently been demonstrated [30].

Here, we demonstrate a multimodal microscope based on a single 10 fs laser source for Fourier-transform (FT)-CARS microscopy, combined with MPEF/SHG/THG imaging. Using a polygonal mirror scanner for a Fourier-domain delay scanner we demonstrate single-shot CARS spectra with <5 $cm^{-1}$ spectral resolution, with a spectral range spanning >1000 $cm^{-1}$ in the vibrational fingerprint region, with a pixel dwell time between 83.3-333.3 µs. As far as we know, this is the fastest sub-10 $cm^{-1}$ resolution broadband CARS hyperspectral microscopy system demonstrated to date. We demonstrate co-localized multimodal imaging of crystals, and explore the current measurement limits of the FT-CARS modality using toluene in a glass micro-capillary, demonstrating >11 fps kilo-pixel chemical images. We also show low-power high-resolution two-photon excited auto-fluorescence (2PEaF) imaging of plant cells.

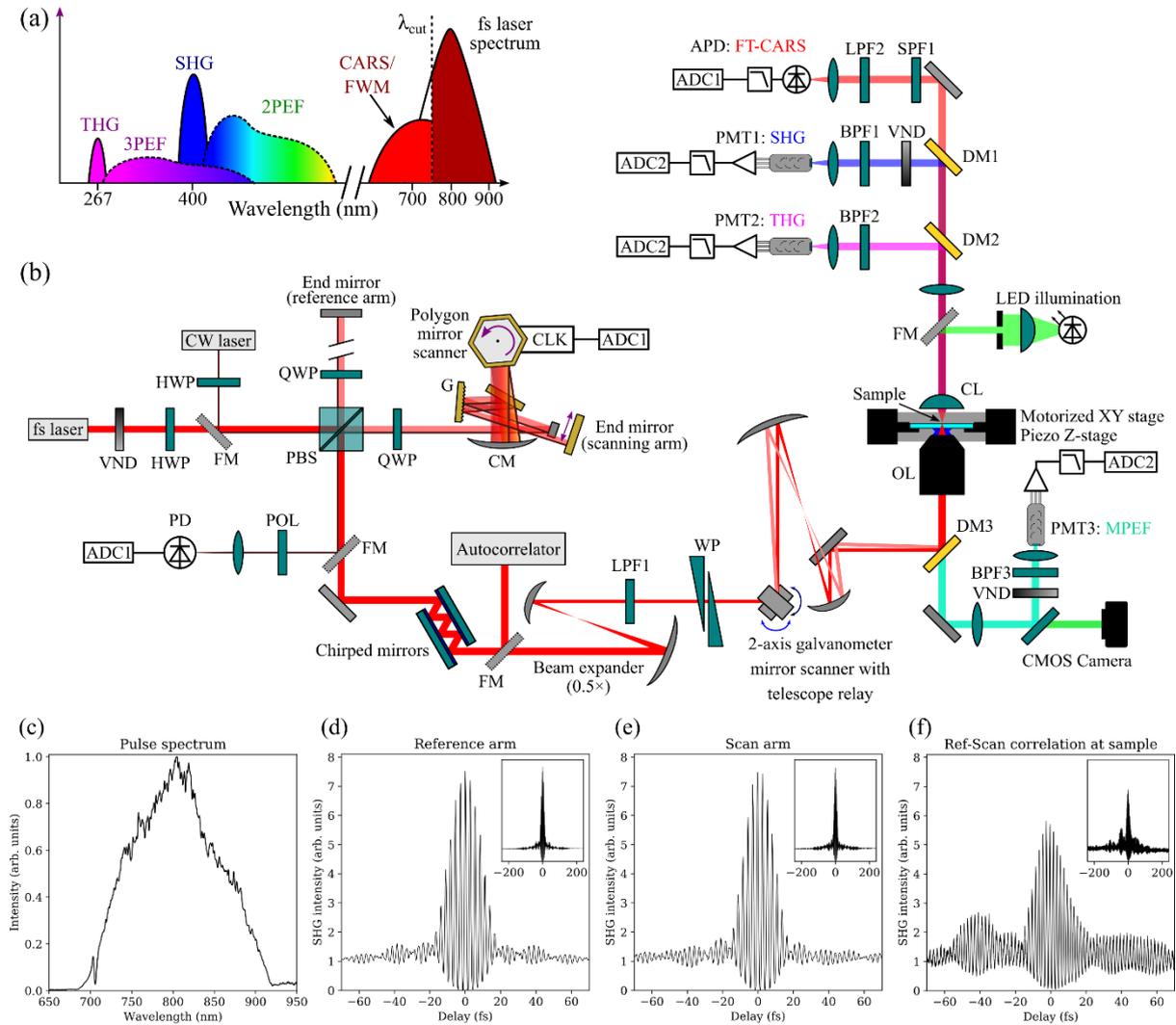

Figure 1. Multimodal FT-CARS microscope. (a) Illustrated spectrum for the optical processes which can be measured with the microscope. $\lambda_{cut}$ refers to the wavelength cut-off selected for the long- and short-pass wavelength filters used for CARS detection. (b) Instrument schematic. VND: Variable Neutral Density filter, HWP: Half-Wave Plate, FM: Flip Mirror, PBS: Polarizing Beamsplitter, QWP: Quarter-Wave Plate, CM: Curved Mirror, G: Diffraction Grating, CLK: polygon scanner clock signal, POL: Linear Polarizer, PD: Silicon Photodiode, LPF: wavelength Long-Pass Filter, WP: Wedge Pair, DM: Dichroic Mirror, OL: Objective Lens, CL: Collection Lens, BPF: Band-Pass Filter, SPF: wavelength Short-Pass Filter, PMT: Photo-Multiplier Tube, APD: Avalanche Photodiode, ADC1: 14-bit, 125 MS/s digitizer board, ADC2 – National Instruments DAQ board. (c) fs laser spectrum. Auto-correlator measurement of reference arm beam (d) and scanning arm beam (e). (f) Cross-correlation of the reference and scan beams at the sample position using a KDP crystal and the SHG detector.

**Methods**

**Instrumentation**

The multimodal FT-CARS instrument shown in Fig. 1(b) utilizes a single mode-locked Ti:Sapphire laser with 10 fs pulse duration, 75 MHz repetition rate, 790 nm center wavelength, 140 nm full-width at half maximum, and 820 mW power output (Synergy PRO, Spectra-Physics). The pulse spectrum at the laser output is shown in Fig. 1(c). The beam passes through a variable ND filter wheel and half-wave plate before entering a polarizing beamsplitter (PBSW-10-800, Optosigma) at the entrance to a

Michelson interferometer. The reference beam passes through a quarter-wave plate and is retro-reflected by a mirror mounted to a motorized stage, which is used for pulse characterization measurements at the sample position. The scanning beam passes through a quarter-wave plate into a folded reflective grating-based 4-f pulse shaper (53066BK02-790R, Richardson Gratings, and CM508-100-P01, Thorlabs) with the polygonal mirror scanner facets placed in the Fourier plane (SA34/DT-18, Lincoln Laser). The dispersed beam is reflected off the polygonal scanner mirror and back through the pulse shaper and on to a retro-reflecting mirror. A 785 nm wavelength continuous wave laser (LPS-785-FC, Thorlabs) is utilized for alignment of the interferometer by placing a mirror after the exit of the interferometer with a linear polarizer and focusing the CW beam onto a photodiode (PDA10A-EC, Thorlabs), and maximizing the visibility of the linear CW interferogram. The reference and scan beams from the femtosecond laser are recombined at the PBS, and pass through a dispersion compensation setup consisting of chirped mirrors (DCMP175, Thorlabs) and a wedge pair (25RB12-01UF.AR2-M, Newport). A beam expander consisting of two curved mirrors is used to resize the beams entering the microscope system. The beams then pass a 700 nm wavelength long-pass filter (FELH0700, Thorlabs) before a galvanometer-scanning mirror pair (6210H, Cambridge Technology), which is relayed towards the microscope using a reflective telescope system. A dichroic filter (FF670-SDi01-25x36, Semrock) directs the laser towards a microscope objective (UPLSAPO 1.2 NA, 60× water immersion or LCPlanN 0.65, 50×, Olympus), which focuses the beams onto a sample mounted on a motorized stage (MLS203-1, Thorlabs) with z-axis piezo-stage (MZS500-E, Thorlabs), which is used for the 3D acquisitions shown in Fig. 3(c) and Fig. 5(a). The SHG, THG and CARS generated at the sample is collected in the forward-direction by a 0.69 NA lens (33-949, Edmund Optics) and subsequently collimated with an additional lens (40-280, Edmund Optics). The signals in the forward direction are then filtered by the various wavelength regions for each processes shown in Fig. 1(a). The THG light between 257-275 nm is reflected by a laser harmonic separator (042-2485, EKSMA), and then passes through a bandpass filter (67-811, Edmund Optics), and roughly focused (33-954, Edmund Optics) onto a photomultiplier tube (PMT) (H11461-09, Hamamatsu) connected to a transimpedance amplifier (TIA60, Thorlabs). The SHG is reflected by a dichroic mirror (FF510-Di02-25x36, Semrock), passing through a variable ND filter, then band-pass filtered (FBH400-40, Thorlabs) and roughly focused (LA1131-A, Thorlabs) onto a PMT (H10720-210, Hamamatsu) with another transimpedance amplifier. The remaining light passes through a 650 nm wavelength long-pass filter (FELH0650, Thorlabs) to reject any residual MPEF/SHG light, and a 700 nm wavelength short-pass filter (FESH0700, Thorlabs) before being focused (LB1471-B, Thorlabs) onto an avalanche photodiode (APD) (PDA10A-EC, Thorlabs). The MPEF light is collected in the backscattering direction with the objective lens, passing through the multiphoton dichroic filter and collimated with a lens (AC254-200-A, Thorlabs). A 50:50 beamsplitter is used to direct the MPEF photons through an interchangeable bandpass filter element (FB550-40, Thorlabs, used in Fig. 5), after which it is roughly focused (AC254-050-A, Thorlabs) onto a PMT (PMT2101/M, Thorlabs). A CMOS camera (DCC1545M, Thorlabs) is also utilized for bright-field microscope imaging, with 510 nm LED illumination. The MPEF, SHG and THG PMT signals are electronically low-pass filtered at 1 MHz (EF508, Thorlabs), and read in via a data acquisition card (National Instruments PCIe-6351) and processed into images in custom LabVIEW software. The CARS APD signal is low-pass filtered at 32 MHz (BLP-30+, Minicircuits) or 48 MHz for the measurements in Fig. 4(a) (BLP-50+, Minicircuits) and acquired using a 4-channel, 14-bit, 125 MS/s PCIe digitizer board (ATS9440, AlazarTech).

The total laser power at the sample position for FT-CARS measurements is typically 200-220 mW, with a 2:1 power ratio between the pump (reference) and probe (scanning) beams. FT-CARS image datasets are currently acquired as single large acquisitions of the APD signal, with the galvano-mirror slow scan control voltage used as a trigger signal. The galvano-mirror fast signal and polygonal scanner clock signal are also acquired as reference timing signals during data processing. For these measurements, the imaging pixel dwell time, $t_d$, is determined by the polygonal mirror facet scan rate ($f_{PM}$, ranging from 3.0-16.5 kHz) as

$$t_d = \frac{1}{f_{PM}} \tag{1}$$

which, for a square image of $N_p \times N_p$ pixels requires that the galvanometer mirrors have a fast axis scanning frequency of

$$f_x = \frac{1}{(2N_p \cdot t_d)} \tag{2}$$

and a slow axis scanning frequency of

$$f_y = \frac{1}{(2N_p^2 \cdot t_d)} = \frac{1}{2t_{im}} \tag{3}$$

for a single-swept forward or backward image of total acquisition time $t_{im}$. The number of samples for the total image can then simply be calculated using the sample rate ($S$, 125 MS/s in current setup) as $N_s = S \cdot t_{im}$. The size of the stored binary data file ($D$) for each digitizer channel can then also be calculated from the bit-depth of the digitizer (currently 14-bit, stored as $B = 2$ bytes per sample) using $D = N_s \cdot B$. As an example for the current system, a single 32×32 pixel image at $f_{PM}$= 3 kHz will result in a raw data trace of ~85.3 MB acquired in $t_{im}$ = 0.341 s

**Pulse measurement**

The individual output beams from each arm of the Michelson interferometer were separately analyzed by an interferometric auto-correlator (Femtometer, Spectra-Physics) after the chirped mirrors in Fig. 1(b). In these measurements the chirped mirror pairs were adjusted to compensate for the dispersion of the polarizing beamsplitter and other optics used in and before the interferometer. The resulting fringe-resolved autocorrelation traces are shown in Fig. 1(d) for the reference beam, and Fig. 1(e) for the scanning beam. Pulse measurement at the sample was carried out using the SHG signal from a small potassium dihydrogen phosphate (KDP) crystal placed on a glass coverslip, measured by the SHG PMT detector. The motorized mirror in the reference arm of the Michelson interferometer was scanned, providing a fringe-resolved cross-correlation measurement between the beams used for FT-CARS measurement, as shown in Fig. 1(f).

**Data Processing**

FT-CARS datasets were processed in Python (Anaconda distribution) using custom scripts which utilized the core scientific packages. Principal component analysis (PCA) was implemented using the scikit-learn package, including the explained variance function to enumerate how much the variance described in the data by each principal component compared to the total variance in the whole dataset. Briefly, for FT-CARS images the one-dimensional time-domain APD signal is cut into intervals matched to the polygon mirror facet scan frequency, which defines the pixel dwell time. The sequence of interferograms are then truncated such that the pulse zero-delay position (shown by a sharp non-resonant spike signal) is close to the edge, but removed from the window used for FFT. The aligned truncated interferograms are then linearly interpolated to allow resampling which corrects for the nonlinear scanning of the group delay by the polygonal mirror, as outlined in previous work [20]. The

CARS spectrum is then retrieved from the absolute value of the FFT for each interferogram in the imaging dataset (after triangular apodization and zero padding). 3D image data for SHG and THG was processed using ImageJ scripting (FIJI distribution).

**Samples**

KDP was prepared from a super-saturated solution of potassium dihydrogen phosphate powder (795488, Sigma Aldrich) in distilled water, heated to near boiling point for 20 minutes until well-mixed, and left to cool and crystallize. Small (<1 mm thickness) fragments of the higher quality KDP crystals were placed onto a glass coverslip for pulse characterization in Fig. 1 and imaging measurements in Fig. 2. Trans-Stilbene powder (139939, Sigma Aldrich) was roughly dispersed onto a glass coverslip for multimodal measurements in Fig. 3. Toluene was pipetted onto a glass micro-capillary element (J5022-11, Hamamatsu) and placed between two glass coverslips for the measurement in Fig. 4. Green onions were obtained from a local supermarket, with single epithelial layers peeled off and placed between two glass coverslips for measurements in Fig. 5.

**Results**

**Multimodal imaging of crystalline samples**

Fig. 2 shows an example imaging measurement of KDP micro-crystals, comprised of SHG, THG and FT-CARS modalities. The bright-field image in Fig. 2(a) shows the region where the multimodal image was taken, with Fig. 2(b) showing the relative vertical extent of the crystal parts in the different regions labelled (i) and (ii). Exploratory analysis of the FT-CARS hyperspectral dataset was carried out using principal components analysis (PCA), the first principal component (PC1) of which is shown in Fig. 2(c). PC1 shows a single dominant feature, assigned to the 914 $cm^{-1}$ phosphate stretching band [31]. Image contrast can be generated either by band area integration of the spectrum at each pixel, as shown in Fig. 2(d) and Fig. 2(e), or using the PC1 coefficients, as shown in Fig. 2(f). The band area image from the 914 $cm^{-1}$ band and the PC1 component both produce comparable image features from region (ii) of the KDP crystal, as the plane in which the laser-scanning imaging measurement was carried out is within the bulk of region (ii). The band area image for the off-resonant spectral region at 960 $cm^{-1}$ has some small residual features from the crystal region, though this is thought to be due to the noise contributions from the increased non-resonant background signal in these regions (which manifests as a constant DC component in the FT-CARS time-domain signal). The combined multimodal measurement in Fig. 2(g) shows THG from the surface of region (i) (but not region (ii), as the imaging plane is not at the surface as shown in Fig. 2(b)), and SHG from both regions. The latter is likely due to the use of sub-20 fs pulses for efficient excitation of SHG from slightly out-of-focus regions of the bulk of the nonlinear KDP crystal.

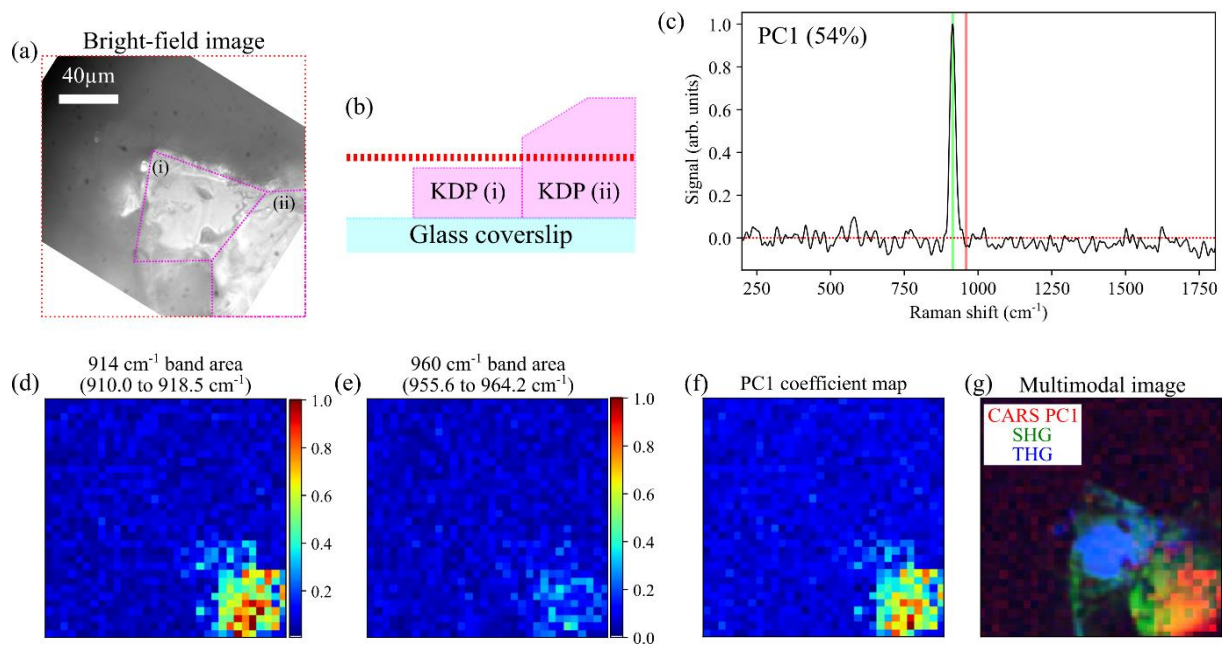

Figure 2. Multimodal measurement of a KDP crystal fragment. (a) Bright-field image of fragment of KDP crystal. The red dashed box shows the region where a co-localized multimodal measurement in of SHG, THG and FT-CARS images was acquired. (b) Schematic showing the estimated vertical profile of the regions of the KDP crystal marked (i) and (ii) in (a), with the red dashed line showing the image plane for multimodal measurement. (c) First principal component for the FT-CARS hyperspectral dataset (54% of explained variance in dataset). The green and red areas show the spectral regions used for the band area images in (d) and (e). (f) FT-CARS image generated using the coefficients of PC1. (g) Combined multimodal image from the SHG, THG and FT-CARS PC1 channels. The FT-CARS image was 32×32 pixels with 333.3 µs pixel dwell time (image acquisition time: 0.34 s), and the SHG and THG images were 100×100 pixels acquired in 0.1 s.

Fig. 3 shows a dataset for a trans-Stilbene (tS) crystal, comprised of SHG, THG and FT-CARS modalities. The 3D SHG/THG measurement in Fig. 3(c) shows the orientation of the top surface of the crystal (from the surface SHG signal in green) with respect to the glass coverslip (surface THG signal in blue). A 100×100 pixel FT-CARS image was acquired in the plane shown near the crystal surface in Fig. 3(b). The mean FT-CARS spectrum of the 100,000 spectra is shown in Fig. 3(d), with the 999 $cm^{-1}$ phenyl ring trigonal deformation, 1194 $cm^{-1}$ C-phenyl stretch, and 1595 $cm^{-1}$ phenyl ring stretching modes observed [32, 33]. The first three PCA components and corresponding coefficient images can be seen in Fig. 3(e). The PC1 component appears similar to the mean spectrum in Fig. 3(d), and the associated coefficient image appears to be highly localized to a region of the tS crystal, particularly when compared to the co-localized SHG frame, as shown in Fig. 3(f). The PC2 component appears to pick out a feature of decreasing signal intensity with increased Raman shift. This might relate to spatial variations in the durations of the excitation pulses (which determine the CARS bandwidth), or possibly some slight variation in intensity due to the cross-polarized excitation relative to the orientation of the local crystal structure (and hence the Raman scattering tensor for each vibrational mode). The latter reasoning may also explain the features in PC3, although detailed analysis of these variations are beyond the scope of this work. Fig. 3(g) shows the positive (red) and negative (blue) PC2 coefficients overlaid onto the SHG image, showing subtler changes than the mean spectrum in the PC1 image.

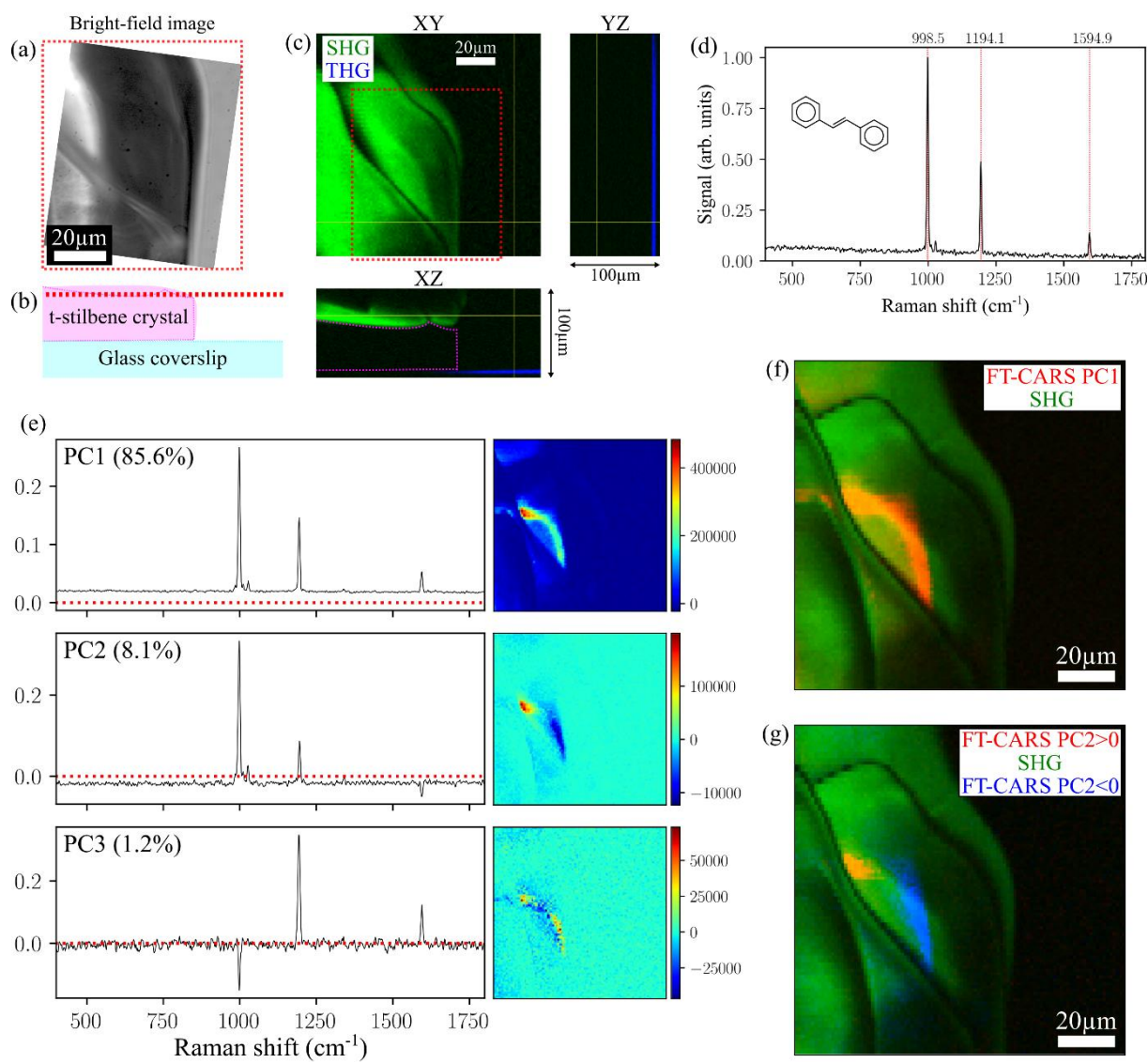

Figure 3. Multimodal measurement of a trans-stilbene crystal. (a) Bright-field image of tS crystal. (b) Schematic of sample in vertical direction, with dashed red line showing estimated plane for FT-CARS image acquisition. (c) Volumetric imaging of tS crystal with SHG (green) and THG (blue) channels. Dashed red box corresponds to the bright-field area shown in (a), while purple dashed line in XZ is a guide for the eye, relating to the schematic (b). (d) Mean spectrum of a 100×100 pixel FT-CARS hyperspectral image dataset acquired near the surface of the crystal with 333.3 µs pixel dwell time (image acquisition time: 3.33 s). (e) Principal component analysis applied to the FT-CARS dataset, showing the vectors and coefficient images for PCs 1-3. (f) Multimodal combined image of FT-CARS PC1 (red) and SHG (green). (g) Another possible multimodal image with FT-CARS PC2>0 (red), SHG (green) and FT-CARS PC2<0 (blue).

**FT-CARS kilopixel imaging at >11 fps**

The polygonal scanner in the current setup can achieve a maximum spectral acquisition rate of 16.5 kHz (60 µs dwell time). However, the down-conversion factor from molecular to RF frequencies is proportional to the polygonal mirror scanning frequency, as shown previously [20]. In the present case, the maximum scan rate places the majority of the Raman fingerprint region above the Nyquist frequency limit, which is currently set by the laser repetition rate as shown in Fig. 4(a). Therefore, the fastest rate currently achievable in the setup whilst maintaining reasonable sampling conditions which avoids aliasing artefacts in the desired measurement region is 12 kHz (83.3 µs). Fig. 4(b) and Fig. 4(c) show measurements of toluene in a glass micro-capillary (10 µm capillary diameter) at different polygonal mirror scanning frequencies. Fig. 4(b) shows a single 32×32 pixel FT-CARS image acquired with a pixel dwell time of 333.3 µs (3 kHz spectral rate), corresponding to an image acquisition time of 0.34 s (~ 3 fps). Using both band area and PCA approaches picks out clear image features attributed to the capillary glass and toluene. PC1 appears to show information relating to the capillary glass, which includes a constant offset value (caused by the increased noise from the non-resonant signal in glass), a broad band around 300 cm$^{-1}$, and negative toluene-associated features, while PC2 closely matches the profile of the toluene Raman spectrum, with the 787 cm$^{-1}$, 1004 cm$^{-1}$, 1031 cm$^{-1}$, 1210 cm$^{-1}$ bands visible. Fig. 4(c) shows a similar 32×32 pixel imaging measurement, but for 27 sequential imaging frames with the pixel dwell time set to 83.3 µs (12 kHz spectral rate), which results in the acquisition time of a single frame in 0.0853 s (~ 11.7 fps). Using PCA to generate images, the glass capillary features can again be seen in PC1 (features slightly offset due to stage jitter), and toluene features again observed in PC2, though without the 1210 cm$^{-1}$ band, likely due to the Nyquist limited cutoff mentioned previously. In this case, the profile of PC1 contains positive features of toluene, suggesting that there may possibly be a layer of toluene between the capillary and glass coverslip, in addition to the toluene within the capillary, as suggested by the mean PC2 image of all sequential frames. Nevertheless, the overlay of the PC1 and PC2 features from the >11 fps measurement agree with the 3 fps measurement in Fig. 4(b).

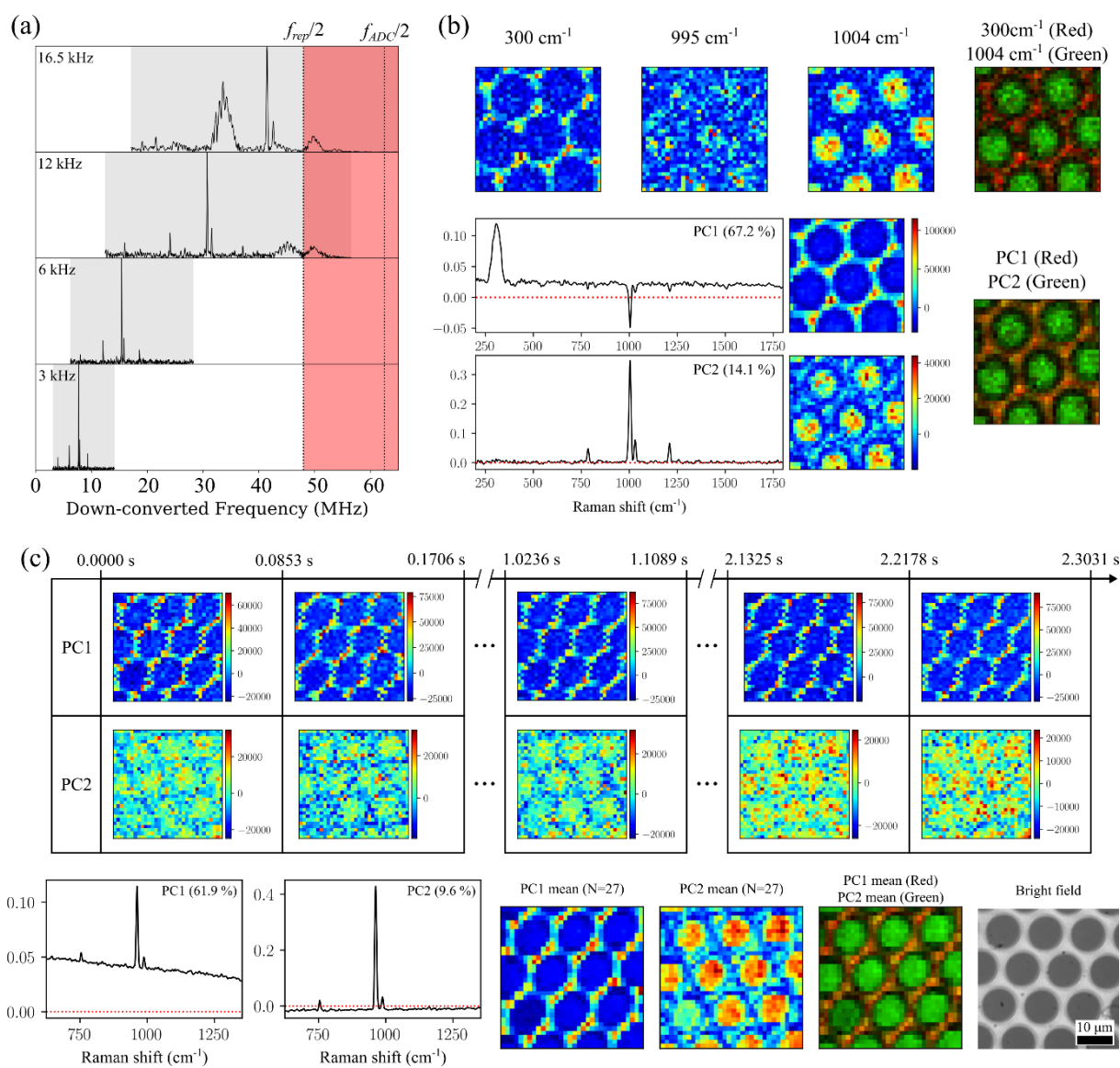

**Figure 4.** High-speed FT-CARS micro-spectroscopy. (a) 100-averaged toluene FT-CARS spectra acquired at different polygonal mirror scanning rates from 3.0-16.5 kHz. Grey shaded regions indicate a Raman bandwidth spanning 400-1800 cm$^{-1}$. Also shown are the Nyquist frequencies corresponding to the laser repetition rate ($f_{rep}/2$) and digitizer ($f_{ADC}/2$). (b) Single frame acquisition at ~3 fps, producing a 32×32-pixel image with pixel dwell time 333.3 µs (3 kHz spectral acquisition rate). Image contrast is generated from the specified band areas or principal component analysis of the 1024 spectra. (c) Continuously acquired frames at >11 fps, with 27 frames of 32×32-pixel images with pixel dwell time 83.3 µs (12 kHz spectral acquisition rate). Image contrast is generated using principal components analysis applied to all 27,648 sequentially acquired FT-CARS spectra. See Visualization 1 for video of all frames.

**Low-power two-photon excited auto-fluorescence bioimaging**

The use of an ultrashort pulse enables more efficient excitation of nonlinear optical interactions, such as multiphoton excited fluorescence. Therefore, the use of a sub-20 fs pulsed excitation can enable low average power imaging, while still generating sufficient photon number for rapid high-resolution imaging. Fig. 5 shows example two-photon excited auto-fluorescence (2PEaF) measurements of a layer of epithelial cells from a green onion, with auto-fluorescence detected between 550±20 nm. Fig.

5(a) shows a 100×100×25 pixel 3D 2PEaF dataset for the cell layer with 0.5 s acquisition for each 2D frame, clearly showing features from the cell walls in the orthogonal projection. The contrast is likely to be generated from a combination of nicotinamide adenine dinucleotide (NAD) and flavins [4]. Fig. 5(b) shows an averaged 2PEaF image from 62 sequential 1000×1000 pixel frames, with Fig. 5(c) showing examples of 5-averaged sequential frames overlaid onto the mean image from Fig. 5(b) (see Visualization 2 for full video). With low power excitation (<1 mW), no damage was apparent in the 2PEaF or bright-field images after several minutes (62 frames of 5 s acquisition time each). However, above ~10 mW excitation, damage was observed during scanning, with rapid micro-bubble formation observed at higher laser power, which agrees approximately with previous reports [34,35].

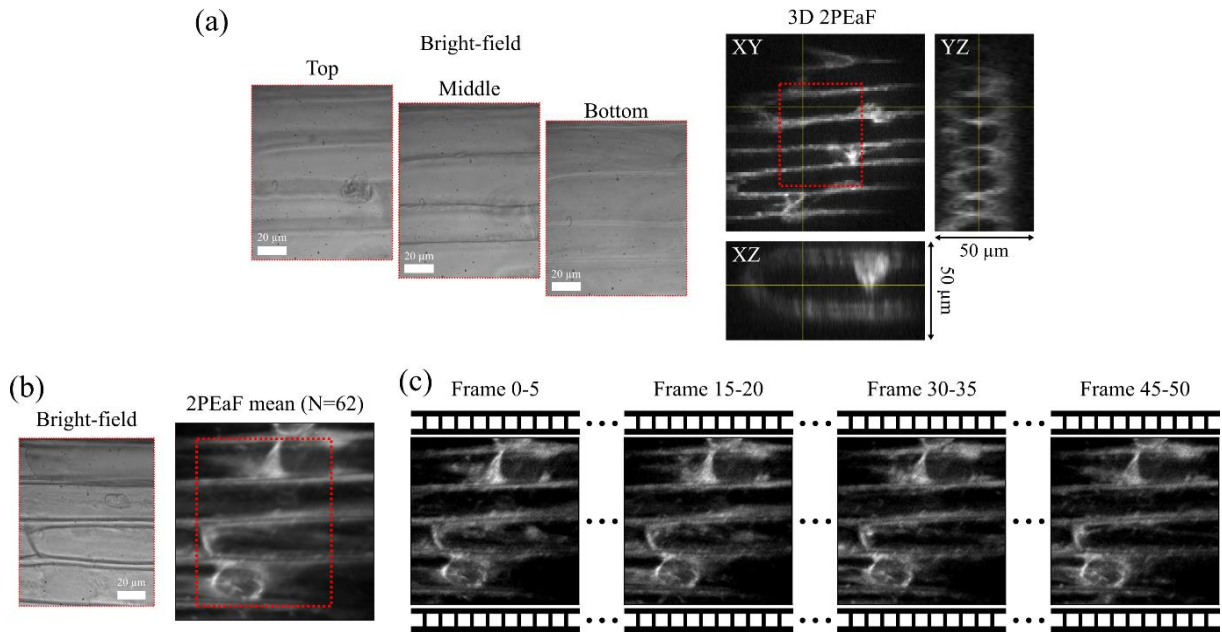

Figure 5. Two-photon excited auto-fluorescence (2PEaF) imaging of onion epithelial cell layer with <1 mW ultrashort pulse excitation (detection between 550±20 nm). (a) 3D 2PEaF measurement of green onion cell monolayer (100×100×25 pixels, 0.5 s acquisition per image slice). (b) Mean measurement of green onion cell layer across 62 frames of 1000×1000 pixel 2PEaF images (5 s acquisition per frame). (c) Example 5-averaged frames for the time-course 2PEaF measurement. See Visualization 2 for full video of data in (c).

**Discussion**

We have demonstrated a multimodal microscope which utilizes a single ultrashort mode-locked laser source for MPEF/SHG/THG imaging and rapid broadband hyperspectral FT-CARS imaging. The FT-CARS modality is capable of high resolution (<5 cm$^{-1}$), broadband (>1000 cm$^{-1}$) micro-spectroscopy at spectral acquisition rates up to 12 kHz, corresponding to pixel dwell times as short as 83.3 µs for imaging. We demonstrated SHG/THG and FT-CARS imaging for KDP and trans-stilbene micro-crystals, >11 fps hyperspectral chemical imaging of toluene in a glass capillary, and low-power multiphoton auto-fluorescence imaging of dynamic processes in plant cells.

The spectral range of the FT-CARS modality is currently limited by the femtosecond pulse width at the sample position, as indicated by the SHG-correlation in Fig. 1(f), which appears to contain features due to third order material dispersion primarily from the high magnification microscope objectives not compensated for by the chirped mirrors. Furthermore, the presence of the 1595 cm$^{-1}$ trans-stilbene band, but absence of the 1639 cm$^{-1}$ band suggests the current upper limit of the spectral range is around 1600-1630 cm$^{-1}$ (corresponding to a vibrational period ~20 fs at the sample). With additional fine-tuning of the dispersion compensation, it should be possible to extend the Raman spectral range up to 3000 cm$^{-1}$ without compromising instrument performance, as recently demonstrated [36]. This could potentially allow simultaneous measurement of the high-wavenumber Raman bands commonly observed in biological samples, as well as Raman labels with strong bands in the silent spectral region. This could be incorporated with spatial light modulator pulse characterization and correction techniques (e.g. multiphoton intrapulse interference phase scanning [37]) to compensate for higher-order dispersion, possibly enabling FT-CARS micro-spectroscopy with almost double the spectral bandwidth demonstrated here. Improved dispersion compensation should also improve the detection sensitivity.

While we were able to demonstrate multiphoton fluorescence bioimaging with low-power femtosecond excitation, the current sensitivity limit for FT-CARS is only suitable for molecules with uncommonly large Raman cross sections, such as carotenoids [22, 38], which are not ubiquitous in biological samples like other bio-macromolecules such as proteins, lipids and nucleic acids. In order to extend FT-CARS towards sensitive bioimaging of such molecules, the photon budget should be allocated efficiently, which may require replacing the polygonal mirror scanner for rapid delay scanning with a resonant scanner and using alternative spectrum estimation approaches to obtain high spectral resolution [39].

The sensitivity should also be improved to allow FT-CARS signal detection at average incident power below ~1-10 mW which are typically used for 2PEF imaging using ultrashort pulses from Ti:Sapphire laser systems [34]. Above this, various mechanisms of sample damage typically occur on microscopic scanning areas, ranging from permeation of cell membranes, to cavitation and formation of micro-bubbles [34, 35]. The detailed study by Hopt and Neher on nonlinear photodamage mechanisms catalogues the relevant parameters which lead to damage [40]. In particular, when only varying the pulse duration, $\tau$, it was found that the time until damage found to be $\propto \tau^{1.5}$ (i.e. shorter pulse, shorter time until damage). As FT-CARS fundamentally requires ultrashort pulse duration, as well as high repetition rate and has a fixed dwell time set by the delay-scanning mechanism, the only option in the current regime of excitation power would be to scan larger areas in shorter time (i.e. decrease spatial resolution). To avoid this, the current excitation power would need to be reduced by 1-2 orders of magnitude to be in the regime of low time-to-damage for microscopic resolution. Future work should therefore focus on enhancing the detection sensitivity to enable FT-CARS detection at such power levels, possibly using heterodyne enhancement from the probe pulse [41], optical subtraction of the pump pulse (background signal) [42], or possibly temporal stretching of the pump and probe pulse to enable lock-in detection [43]. The latter two approaches may also be beneficial in terms of enhancing the dynamic range of measurement, which is currently limited as the AC resonant-CARS signal of interest is measured on top of a large DC non-resonant component.

Achieving the goals outlined above should allow a setup similar to that demonstrated here to be capable of sensitive non-invasive multimodal nonlinear imaging of cells and tissues.

**Funding**

F.S. acknowledges financial support from the Japanese Society for Promotion of Science (JSPS). This work was financially supported by the New Energy and Industrial Technology Development Organization (NEDO) project (P16011), JST PRESTO (JPMJPR17G2), and JSPS KAKENHI (17H04852, 17K19071, 18F18012).

**Acknowledgements**

The authors acknowledge assistance with instrument construction from Shunya Konno and Hidekazu Shirai, and thank Prof. J. Yumoto and Prof. M. Gonokami for use of their equipment.

**Disclosures**

The authors declare no conflicts of interest.